\documentclass[preprint2]{aastex}

\usepackage{graphicx}
\usepackage{graphics}
\usepackage{txfonts}
\usepackage{epsfig}
\usepackage{natbib}
\usepackage{times}
\usepackage{amssymb}
\usepackage{color}

\newcommand\igr{\object{IGR\,J16207$-$5129}}
\newcommand\ergcms{erg\,cm$^{-2}$\,s$^{-1}$}
\newcommand\ergs{erg\,s$^{-1}$}
\newcommand\integ{{\it{INTEGRAL}}}
\newcommand\suz{{\it{Suzaku}}}

\newcommand\swift{{\it{Swift}}}
\newcommand\xmm{{\it{XMM-Newton}}}
\newcommand\chan{{\it{Chandra}}}
\newcommand\nh{$N_\mathrm{H}$}
\definecolor{red}{rgb}{0.7,0,0}
\definecolor{blue}{rgb}{0,0,0.7}

\shorttitle{Broadband \suz\ observations of \igr}
\shortauthors{Bodaghee et al.}

\begin{document}

\title{Broadband \suz\ observations of \igr}

\author{A. Bodaghee and J. A. Tomsick}
\affil{Space Sciences Laboratory, 7 Gauss Way, University of California, Berkeley, CA 94720, USA}
\email{bodaghee@ssl.berkeley.edu}

\and

\author{J. Rodriguez and S. Chaty}
\affil{Laboratoire AIM, CEA/IRFU - Universit\'e Paris Diderot - CNRS/INSU, \\ CEA DSM/IRFU/SAp, Centre de Saclay, F-91191 Gif-sur-Yvette, France}

\and

\author{K. Pottschmidt}
\affil{CRESST and NASA Goddard Space Flight Center, Astrophysics Science
Division, Code 661, Greenbelt, MD 20771, USA \\
Center for Space Science and Technology, University of Maryland Baltimore
County, 1000 Hilltop Circle, Baltimore, MD 21250, USA}

\and

\author{R. Walter}
\affil{\integ\ Science Data Centre, Universit\'e de Gen\`eve, Chemin d'Ecogia 16, CH--1290 Versoix, Switzerland \\ 
	Observatoire de Gen\`eve, Universit\'e de Gen\`eve, Chemin des Maillettes 51, CH--1290 Sauverny, Switzerland}

\begin{abstract}
An analysis of \igr\ is presented based on observations taken with \suz. The data set represents $\sim$80\,ks of effective exposure time in a broad energy range between 0.5 and 60\,keV, including unprecedented spectral sensitivity above 15\,keV. The average source spectrum is well described by an absorbed power law in which we measured a large intrinsic absorption of \nh\ $= (16.2_{-1.1}^{+0.9}) \times 10^{22}$\,cm$^{-2}$. This confirms that \igr\ belongs to the class of absorbed HMXBs. We were able to constrain the cutoff energy at $19_{-4}^{+8}$\,keV which argues in favor of a neutron star as the primary. Our observation includes an epoch in which the source count rate is compatible with no flux suggesting a possible eclipse. We discuss the nature of this source in light of these and of other recent results.
\end{abstract}

\keywords{accretion, accretion disks ; gamma-rays: general ; stars: neutron ; supergiants ; X-rays: binaries ; X-rays: individual (\igr)  }

\section{Introduction}

%
\begin{deluxetable}{ l c c c c }
\tablewidth{0pt}
\tablecaption{Journal of Suzaku observations of IGR\,J16207$-$5129 \label{tab_log}}
\tablehead{
\colhead{observation ID}      	& 
\colhead{dates}				& 
\colhead{instrument}  		&
\colhead{exposure time} 		&
\colhead{name}           		\\ 
\colhead{}      				& 
\colhead{(MJD)}			& 
\colhead{}  				&
\colhead{(ks)}				&
\colhead{}      				
}
\startdata
	402065010 	& 54499.823--54500.784	& HXD			& 49.596	 & O1 \\
	402065020 	& 54526.866--54527.750 	& HXD			& 28.331   & O2 \\
				& 					& XIS		 	& 32.678	 & 	 \\
\enddata
\vspace{-7mm}
\end{deluxetable}

Seven years after its launch, \integ\ \citep{Win03} continues to discover soft gamma-ray sources in the 20--100\,keV energy range \citep{Bir10}. So-called ``IGRs'' (for \integ\ Gamma-Ray sources\footnote{for an updated list of IGRs, please visit http://irfu.cea.fr/Sap/IGR-Sources/}) now number over 300. While \integ's gamma-ray imager ISGRI \citep{Leb03} discovered these objects, it provides an error radius that is typically around 2--4$^{\prime}$, i.e. too large to allow for the identification of an optical/IR counterpart which would help classify the source into one of the many families of high-energy emitters. Therefore, follow-up observations with focusing X-ray telescopes such as \chan\ \citep[e.g.][]{Tom06}, \swift\ \citep[e.g.][]{Rod10}, and \xmm\ \citep[e.g.][]{Bod06} are necessary to refine the X-ray position and to extend the spectrum below ISGRI's $\sim$20\,keV lower limit. 

One such source, \igr, was discovered in the first mosaic images constructed from \integ-ISGRI scans of the Galactic Plane \citep{Wal04,Tom04}. A follow-up observation with \chan\ allowed \citet{Tom06} to refine the source position to within an error radius of 0.6$^{\prime\prime}$ which excludes all but a single optical/IR counterpart (\object{2MASS J16204627$-$5130060} = \object{USNO-B1.0 0384-0560875}) whose spectral energy distribution is that of a hot and massive late O or early B-type star \citep[e.g.][]{Tom06,Mas06c,Neg07,Rah08,Nes08}. The system is therefore classified as a supergiant HMXB (SGXB).

The X-ray characteristics of \igr\ lend further credence to this classification. Although its emission shows variability by at least an order of magnitude, \igr\ is persistent having been detected in short and long-exposure images taken at various times with \integ\ \citep{Bir04,Bir06,Bir07,Bir10,Kri07} and other X-ray telescopes \citep[e.g.][]{Tom06,Mas06c,Tom09}. An absorbed power law fits the 0.3--10\,keV \chan\ spectrum well with $\Gamma = 0.5_{-0.5}^{+0.6}$ \citep{Tom06}. In the \integ\ 20--50-keV band, the spectral slope steepens to $\Gamma = 1.9\pm0.5$ possibly indicating a high-energy cutoff \citep{Tom06}. Using \xmm, \citet{Tom09} measured a column density of \nh\ $ = (11.9_{-0.5}^{+0.6})\times10^{22}$\,cm$^{-2}$. This large intrinsic column density firmly establishes \igr\ among the growing class of absorbed SGXBs. Other spectral features typical of this class were unveiled such as a K$\alpha$-emission line at $6.39\pm0.03$\,keV from iron fluorescence, and soft excess emission at $\sim2$\,keV consistent with the reprocessing of X-rays in the wind \citep{Tom09}. 

\suz\ is playing an increasing role in providing spectral and timing follow-up observations of IGRs \citep[e.g.][]{Boz08,Mor09,Pai09}. Recently, we asked \suz\ to target \igr. Here, we present the results from the analysis of these data including: an X-ray spectrum in the 0.5--10\,keV range that confirms the obscured environment of the X-ray source; a spectrum with unprecedented sensitivity between 15 and 60\,keV allowing us to constrain the cutoff energy; and a light curve that includes a possible eclipse. The key steps of the analysis are presented in Section\,\ref{obs}, followed by the pertinent results from timing and spectral studies in Section\,\ref{res}. Our conclusions and the implications of our findings are discussed in Section\,\ref{disc}.

\section{Observations}
\label{obs}

Launched in 2005, the \suz\ space telescope features 2 co-aligned instruments working in tandem to observe the hard X-ray sky (0.2--600\,keV): the X-ray Imaging Spectrometer \citep[XIS:][]{Koy07} and the Hard X-ray Detector \citep[HXD:][]{Tak07}. XIS consists of 2 front-illuminated CCD cameras (XIS0 and XIS3), and one that is back-illuminated (XIS1). HXD contains 2 detectors, PIN and GSO, which are mainly sensitive to energies below 60\,keV and above 40\,keV, respectively.

\begin{figure*}[!t] 
\centering
\includegraphics[width=16cm,angle=0]{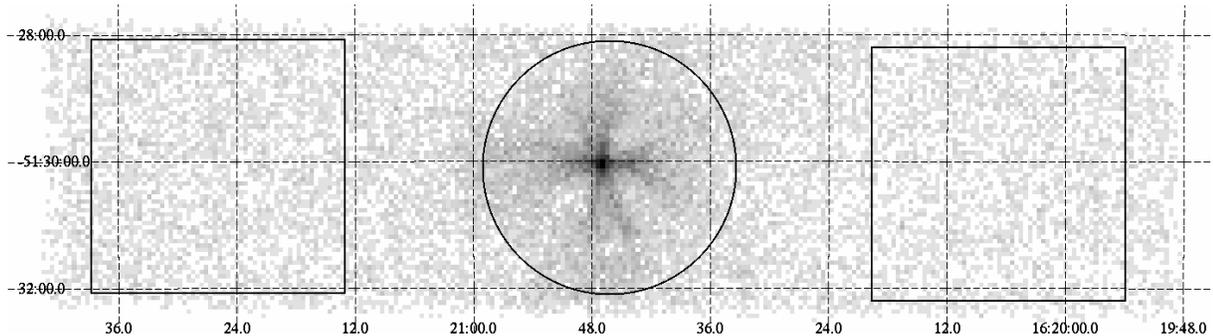}
\caption{Image of IGR\,J16207$-$5129 as captured with the XIS1 detector on {\it Suzaku}. Pixels have been grouped in blocks of 4 (bin: group4 in \texttt{ds9}). The circular source and squared background selection regions are also shown. }
\label{img}
\end{figure*}

\begin{figure}[!t] 
\centering
\includegraphics[width=7cm,angle=0]{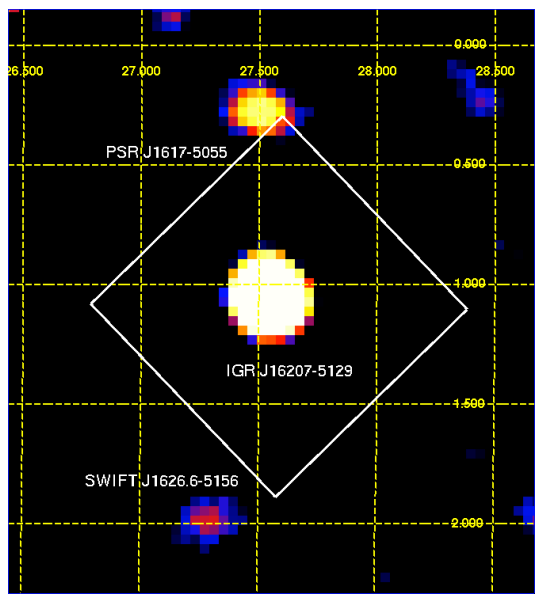}
\caption{Significance image (from 3 to 10$\sigma$) in the 17--80 keV band of IGR\,J16207$-$5129 from {\it INTEGRAL}-ISGRI. The image is in Galactic coordinates and the white box represents the FOV from HXD-PIN. }
\label{isgimg}
\end{figure}

Focused on \igr, observation ID 402065010 (PI: J. Tomsick) took place on February 3--4, 2008. A technical glitch affected the on-board processor for XIS resulting in no data from the low-energy instrument. Our observing time was rescheduled for March 1--2, 2008, as observation ID 402065020. This time, both instruments recorded data successfully. All observations were performed in the XIS-nominal pointing mode while using a quarter window option in order to gain a higher time resolution. Table \ref{tab_log} presents the observation log.

Data were reduced with v.6.7 of the HEAsoft software package following the procedures described in the \suz\ ABC Guide v.2. The task \texttt{xispi} converted the unfiltered XIS event files to pulse-invariant channels, which were then input into \texttt{xisrepro} to return cleaned event files. After removing bad events, we extracted the source spectrum from each XIS detector using a circular region of 120$^{\prime\prime}$-radius centered on the position given by \chan\ (see Fig.\,\ref{img}). A pair of 240$^{\prime\prime}$-wide squares positioned on a source-free region of the detector served as the background. Response matrices were generated with \texttt{xisrmfgen} and \texttt{xissimarfgen} taking into account these source and background extraction regions. To improve photon statistics, we merged the source spectral files for the front-illuminated CCDs (XIS0 and XIS3), and we did the same for the background. A minimum of 150 counts was allocated to each bin in the XIS spectra. 

We used the tuned (v.2.0) background file and the common GTI during spectral extraction of HXD-PIN. We then performed dead-time corrections with \texttt{hxddtcor}. We modeled the cosmic X-ray background using the flat-field response file from January 29, 2008, and then subtracted its contribution from the HXD-PIN background spectrum. We tested whether the HXD-PIN spectra from O1 and O2 were consistent in terms of power-law slope ($\Gamma \sim 2$) and normalization before they were summed into a single spectrum. Channels in this spectrum were paired up between 15 and 20\,keV, grouped by fours between 20 and 40\,keV, and collected in a single bin between 40 and 60\,keV. Spectral bins above this were excluded.

\begin{figure*}[!t] \centering
\includegraphics[width=16cm,angle=0]{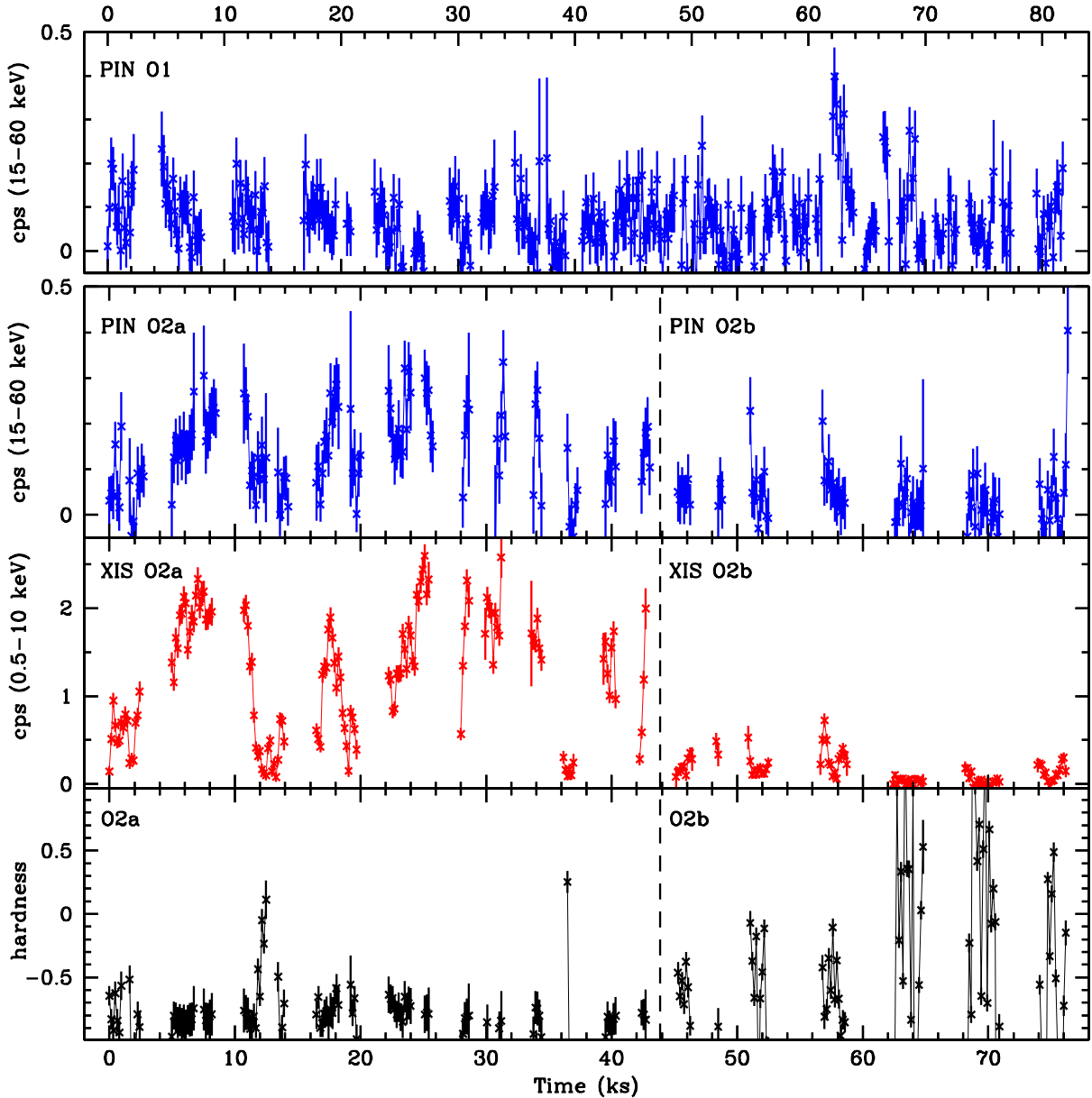}
\vspace{-3mm}
\caption{Background-subtracted light curve of IGR\,J16207$-$5129 from XIS (red: 0.5--10\,keV) and HXD (blue: 15--60\,keV). The upper panel displays observation O1 while the middle and lower panels show O2 (see Table\,\ref{tab_log} for start times). Each bin collects 160\,s worth of data. The hardness ratio is defined as $\frac{H-S}{H+S}$, where $H$ and $S$ are the count rates in 0.5--10\,keV and 15--60\,keV, respectively. The dashed line represents MJD\,54527.382 which corresponds to the onset of a period of low activity. }
\label{lc_x0_hxd}
\vspace{-4mm}
\end{figure*}

Source and background light curves were generated in 2 energy ranges, 0.5--10\,keV (XIS) and 15--60\,keV (HXD-PIN), for an effective exposure time of $\sim$80\,ks. The time resolution was set to 2\,s for XIS, and 32 and 64\,s for HXD-PIN. 

Figure\,\ref{isgimg} shows that the millisecond pulsar named \object{PSR\,J1617$-$5055} is near the edge of the nominal HXD-PIN field of view (FOV). In the ISGRI energy band in which this image was captured (17--80\,keV), the ms-pulsar has an average flux around a quarter that of \igr. According to the \suz\ technical description, the HXD-PIN count rate from a Crab-like source would decrease by a factor of 20 when offset at 30$^{\prime}$ from the center of the FOV. This implies that the contribution expected from this source to the \igr\ count rate is 1--2\% on average, and up to 10\% if the flux from \igr\ diminishes by an order of magnitude.

\begin{figure}[!t] \centering
\includegraphics[width=5.5cm,angle=-90]{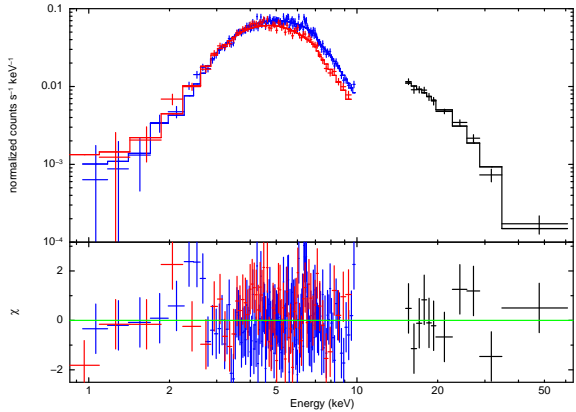}
\caption{Full spectrum of IGR\,J16207$-$5129 corrected for the background and fit with an absorbed power law with an exponential cutoff whose parameters are listed in Table\,\ref{tab_spec}. The data represent photon counts from XIS1 (red), XIS0 combined with XIS3 (blue), and HXD-PIN (black). }
\label{spec}
\end{figure}
%

\begin{figure}[!t] \centering
\includegraphics[width=7.5cm,angle=0]{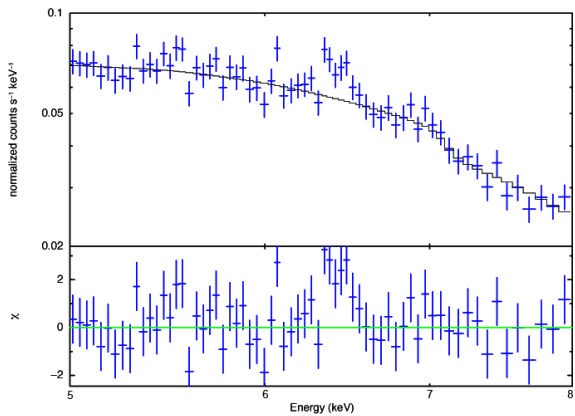}
\caption{Combined XIS0 and XIS3 spectrum of IGR\,J16207$-$5129 in the 5--8\,keV energy range. Each bin collects 150 counts. An absorbed power law was fit to the data (solid curve). An iron line at 6.4\,keV is suggested by the residuals to the model displayed in the bottom panel. }
\label{spec_fe}
\end{figure}
%

\section{Results}
\label{res}

\subsection{Timing Analysis}

Figure\,\ref{lc_x0_hxd} presents the background-subtracted light curves in 160-s bins from both soft and hard instruments. The figure shows that the emission from \igr\ varies within 1--2 orders of magnitude on ks-timescales, i.e. within the range of variability expected of persistent SGXBs \citep{Wal07}. In contrast, supergiant fast X-ray transients \citep[SFXTs: e.g.][]{Neg06,Sgu06} spend most of their time in a quiescent or intermediate phase ($L_{X} \sim 10^{32}$--$10^{33}$\,\ergs) punctuated by flaring episodes lasting a few hours that represent variations of 2--5 orders of magnitude \citep{Sid08}. The average (unabsorbed) flux during the observation is $\sim 3\times 10^{-11}$\,\ergcms\ in the 0.5--10\,keV range which is similar to the average flux found with \xmm\ \citep{Tom09}. Adopting a distance of 6\,kpc \citep{Nes08}, the luminosity recorded in XIS1 at the highest point is $5\times10^{35}$\,\ergs\ (between 24 and 26\,ks after the start of O2), while it is $2\times10^{34}$\,\ergs\ at its lowest point (between 62 and 72\,ks after the start of O2).

Intriguingly, X-rays from the source, and the variability associated with that emission, are significantly diminished during the last $\sim$30\,ks of the observation. The flux, notably in the soft band, is consistently below the average registered for the full O2 data set beginning at around 45\,ks into O2, which corresponds to MJD 54527.382. The light curves for O2 were then accordingly split into the epochs O2a and O2b before and after this date. 

%
\begin{deluxetable}{ l l l l l  l l c l }
\tabletypesize{\scriptsize}
\tablecolumns{9} 
\tablewidth{0pc} 
\tablecaption{Spectral parameters from models fit to XIS 0--3 and HXD-PIN spectra of IGR\,J16207$-$5129 \label{tab_spec}}
\tablehead{
\colhead{\texttt{Xspec} model}				& 
\colhead{$C$}							& 
\colhead{$N_\mathrm{H}$}				& 
\colhead{$\Gamma$}					&
\colhead{$E_{\mathrm{cut}}$}				&
\colhead{$E_{\mathrm{fold}}$}				&
\colhead{$E\left(\mathrm{K}\alpha\right)$}		&   
\colhead{norm.}							&          		
\colhead{$\chi_{\nu}^{2}$/dof}				\\   	
\colhead{}								& 	
\colhead{}								& 
\colhead{$10^{22}$\,cm$^{-2}$}				& 
\colhead{}								&
\colhead{keV}							&
\colhead{keV}							&
\colhead{keV}							&
\colhead{$10^{-3}$}						&
\colhead{}   			
}
\startdata
\cutinhead{O1+O2} 
\vspace{1mm}
c*phabs*pow				& $0.70_{-0.06}^{+0.07}$	& $19.1_{-0.8}^{+0.8}$	& $1.36_{-0.07}^{+0.07}$	& 	& 	& 	& 4.4	 & 1.38/204	\\
\vspace{1mm}
c*phabs*highecut*pow		& $1.05_{-0.14}^{+0.15}$	& $16.8_{-0.8}^{+0.7}$	& $1.09_{-0.09}^{+0.11}$	& $6.4_{-0.5}^{+0.4}$	& $20_{-4}^{+8}$	&	& 2.7 & 1.15/202	\\
c*phabs*highecut*(pow+gauss)	& $0.96_{-0.11}^{+0.13}$	& $16.2_{-1.1}^{+0.9}$	& $0.92_{-0.17}^{+0.24}$	& $4.5_{-0.5}^{+2.6}$	& $19_{-5}^{+6}$	&  $6.42_{-0.03}^{+0.03}$ & 2.1 & 1.05/199	\\
\cutinhead{O2}
\vspace{-1mm}
c*phabs*pow					& $0.73_{-0.08}^{+0.08}$	& $18.7_{-1.0}^{+1.0}$		& $1.32_{-0.08}^{+0.08}$	& 	& 	& 	& 4.1 & 1.30/212 \\
\cutinhead{O2a}
\vspace{-1mm}
c*phabs*pow					& $0.75_{-0.09}^{+0.10}$	& $19.1_{-1.0}^{+1.0}$		& $1.34_{-0.09}^{+0.09}$	& 	& 	& 	& 6.2 & 0.91/182 \\
\cutinhead{O2b} 
c*phabs*pow					& $1.21_{-0.56}^{+1.02}$	& $19.3_{-3.7}^{+4.8}$	& $1.48_{-0.38}^{+0.44}$	& 	& 	& 	& 1.4 & 0.51/64	\\
\enddata
\vspace{-6mm}
\tablecomments{Errors are quoted at 90\% confidence. $C$ represents an instrumental cross-calibration coefficient which is fixed to 1 for XIS and left free to vary for HXD-PIN.}
\vspace{-6mm}
\end{deluxetable}

For XIS, the median flux in O2a is 1.32\,cps (std. dev.: $\sigma=0.7$). This is higher by at least a factor of 10 than the median flux in O2b: 0.11\,cps ($\sigma=0.15$). This difference between O2a and O2b is also reflected in higher orders of the distributions of light-curve bins: O2a was less peaked and more symmetric in its distribution (kurtosis $k=-1.2$, skewness $s=-0.1$) than was O2b ($k=2.2$, $s=1.3$). The HXD light curves also show evidence of a prolonged attenuation (by a factor of 5) of photon count rates during O2b, with clear statistical differences in the higher orders of the distribution. The distribution of light curve bins from HXD-PIN displayed a sharper peak with more symmetry during O2a ($k=-0.6$, $s=0.03$) than during O2b ($k=9.8$, $s=2.1$). 

We searched the light curves from XIS and HXD for periodicities on the order of a few tens of seconds to a few thousand seconds. A moderate signal appears at $\sim$970\,s in the soft and hard bands (significance $\gtrsim 5 \sigma$). While this is in the range of reported spin frequencies for X-ray pulsars accreting from the winds of supergiant companions \citep[][and references therein]{Liu06,Bod07}, this also coincides with a sixth of the orbital period of the \suz\ satellite. The 970-s signal is also present in the background light curve from HXD-PIN, as are fractional harmonics corresponding to a fourth, and a third of the satellite's orbital period. We conclude that this periodicity is due to aliasing of the satellite's orbit. It is worth noting that no periodic signal was detected in a continuous $\sim$50-ks observation with \xmm\ \citep{Tom09}.

\subsection{Spectral Analysis}
\label{sec_spec}

Figure\,\ref{spec} presents the 0.5--60\,keV spectrum of \igr\ from all observations (O1 and O2 combined). The source spectrum was initially fit with a power law that includes a photoelectric absorption component with \citet{Wil00} abundances, and an instrumental cross-calibration constant which was fixed at 1 for XIS, and left free for HXD-PIN. This fit gave a reduced $\chi^{2}$ value of 1.38 for 204 degrees of freedom (dof). Significant residuals remained above 30\,keV, so we added an exponential cutoff which improved the quality of the fit ($\chi_{\nu}^{2}=1.15$ for 202 dof). The optimal fit ($\chi_{\nu}^{2}=1.05$ for 199 dof) was obtained by complementing this cutoff power-law model with a Gaussian line to account for Fe\,K$\alpha$ emission at 6.4\,keV. We expect to find this line in our spectrum since it was detected in a previous observation with \xmm\ \citep{Tom09}. The residuals from the model without an iron line are shown in Fig.\,\ref{spec_fe}.

Parameters for these models are listed in Table\,\ref{tab_spec}, but we cite here specifically the column density of \nh\ $= (16.2_{-1.1}^{+0.9}) \times 10^{22}$\,cm$^{-2}$, $\Gamma = 0.92_{-0.17}^{+0.24}$, and the cutoff energy at $19_{-5}^{+6}$\,keV. Assuming a distance of 6\,kpc \citep{Nes08}, the unabsorbed source luminosity is $3\times10^{35}$\,\ergs\ in the 0.5--10\,keV range, which is twice the average luminosity measured with \xmm\ \citep{Tom09}. 

Visual inspection of the unbinned HXD-PIN spectrum shows residuals to the power-law fit at $\sim$22\,keV, although there are no harmonically-spaced lines (specifically, at twice the energy of the purported fundamental). We considered the possibility of a cyclotron line by adding a Gaussian absorption component (\texttt{gabs} in \texttt{Xspec}). This did not improve the quality of the fit compared to a power law. Furthermore, we could not simultaneously constrain the related optical depth, the line width, and line energy. The null hypothesis can not be excluded given an F-test probability approaching 80\%. Maintaining the width of line at 3\,keV (a typical value found in the literature for cyclotron lines in other sources), we fixed the line energy to integer values ranging from 20 to 30\,keV, which we used to determine an upper limit of $<$3.4 on the optical depth.

We sought to compare the spectrum of the source before and after MJD 54527.382 which corresponds to the onset of a noticeable change in the emission behavior. Therefore, we split the XIS 0--3 and HXD-PIN spectra according to epochs O2a and O2b. Although the HXD-PIN recorded few photon counts during O2b, a careful grouping of the data enabled us to collect 4 usable bins between 15 and 60 keV. We then fit both spectra with absorbed power laws. The power-law slope and column density were consistent in both epochs: $\Gamma = 1.34_{-0.09}^{+0.09}$ and \nh\ $= (19_{-1}^{+1}) \times 10^{22}$\,cm$^{-2}$ for O2a; and $\Gamma = 1.48_{-0.38}^{+0.44}$ and \nh\ $= (19_{-4}^{+5}) \times 10^{22}$\,cm$^{-2}$ for O2b. The only significant difference between the spectra from O2a and O2b was in the normalization.

Besides O2b, there were other instances in which the source count rate in XIS was low or close to zero, such as e.g. around 12\,ks and 36\,ks into O2a (see Fig.\,\ref{lc_x0_hxd}). Therefore, we separated the XIS spectra into low and high states corresponding to source count rates (prior to background subtraction) below and above 0.5\,cps, respectively. The column density from a power-law fit was slightly larger during the low state (\nh\ $= (22.2_{-1.7}^{+1.8}) \times 10^{22}$\,cm$^{-2}$), but remained statistically compatible with the high state (\nh\ $= (19.4_{-1.2}^{+1.2}) \times 10^{22}$\,cm$^{-2}$). Once again, normalization of the spectral model remained the only significant difference between the low and high states.

Defining the hardness ratio as $\frac{H-S}{H+S}$, where $S$ is the count rate in 0.5--10\,keV, and $H$ is the count rate in 15--60\,keV, we find  that the source spectrum was soft at $-1$ during most of O2a, punctuated by 2 positive spikes during the brief episodes of low intensity at 12 and 36\,ks into O2. During O2b, the ratio tended to harder values, but since the dispersion is so large, we can not confidently state that the increase is significant. We redefine the hardness ratio, where $S$ is now the count rate in 0.5--5\,keV, and $H$ is the count rate between 5 and 10\,keV. The boundary at 5\,keV effectively splits the light curve into an equivalent number of total counts, and ensures that variations in the column density will mainly affect the $S$-band. In this case, the hardness ratio during O2a varied between 0.5 and 1.0 with hard spikes during the 2 brief low-flux episodes, and tended to be harder in O2b, but, once again, the uncertainties are large.

According to the hardness ratios, the high-energy X-ray emission appears to be less affected than the low-energy emission during the low-intensity states of O2a and during O2b. The nearby millisecond pulsar \object{PSR\,J1617$-$5055} contributes 10\% of the HXD-PIN emission during O2b which is not enough to account for the observed change in the hardness ratio. We point out that the potential hardening observed in \igr\ during O2b is similar to the ingress phase of the eclipsing SGXB \object{4U\,1538$-$52} in which a non-negligible hard X-ray flux continues to be detected for 20\,ks after the low-energy bandpass has reached a minimum count rate \citep{Rob01}. 

\section{Discussion}
\label{disc}

\subsection{The compact object and its environment}

The photoelectric absorption that we measured for \igr\ is \nh\ $= (16.2_{-1.1}^{+0.9}) \times 10^{22}$\,cm$^{-2}$), which is 12--60\% larger than the value found with \xmm\ \citep{Tom09}, and 3--7 times larger than the value reported with \chan\ \citep{Tom06}, suggesting a variable column density. The expected line-of-sight absorption ($N_{\mathrm{H}}^{\mathrm{G}}$) is among the highest in the Galaxy at (1.6--1.7)$\times10^{22}$\,cm$^{-2}$ \citep[][]{Dic90,Kal05}. Subsequently, we infer that in addition to the interstellar medium that absorbs some of the radiation, there is matter obscuring the X-ray source that is local to the system. This situation is similar to, although not as extreme as, the case of \object{IGR\,J16318$-$4848} \citep[e.g.][]{Fil04}. 

We can estimate the density of material around SGXBs such as \igr\ by integrating the wind density along the line of sight to the X-ray source \citep{Cas75,Lev04,Pra08}:

\begin{equation}
\rho_{w} = \frac{\dot{M_{\star}}}{ 4\pi r^{2} v_{w}}
\label{eq_rho}
\end{equation}

\begin{equation}
v_{w}(r) = v_{\infty} \left (1 - \frac{R_{\star}}{r} \right ) ^{\beta}
\end{equation}

Here, $R_{\star}$ is the radius of the stellar companion, $v_{\infty}$ is the terminal velocity of the stellar wind, $\dot{M_{\star}}$ is the stellar mass-loss rate, and $\beta$ is the wind velocity gradient ($\approx$0.8--1.2, we will assume $\beta=1$). We adopt standard values for B1\,Ia supergiant stars as provided by \citet{Cro06}, specifically $v_{\infty} = 725$\,km\,s$^{-1}$, and $\dot{M_{\star}}=10^{-6}$\,$M_{\odot}$\,y$^{-1}$. The radius of the companion star is set to $R_{\star} = 20$\,$R_{\odot}$. The compact object (CO) is considered to be a neutron star at the canonical mass of 1.4\,$M_{\odot}$. Circular orbits are assumed with $r = 2 R_{\star}$. The binary is viewed edge-on, and the CO is at inferior conjunction. We also introduce the mass of the hydrogen atom ($m_{\mathrm{H}}$) and the mean molecular weight of the accreted matter ($\mu \sim 0.6$ for Solar material). 

Integrating Eqn.\,\ref{eq_rho}, we obtain:
\begin{eqnarray}
N_{\mathrm{H}} & = & N_{\mathrm{H}}^{\mathrm{G}} + \int_{r=2R_{\star}}^{\infty} \rho_{w} dr \\
			 & = & N_{\mathrm{H}}^{\mathrm{G}} + \frac{\dot{M_{\star}}}{ 4\pi \mu m_{\mathrm{H}} v_{\infty}} \int_{2R_{\star}}^{\infty} \frac{dr}{ r^{2} \left (1 - \frac{R_{\star}}{r} \right ) } \\
			 & = & N_{\mathrm{H}}^{\mathrm{G}} + \frac{\dot{M_{\star}}}{ 4\pi \mu m_{\mathrm{H}} v_{\infty}} \frac{\ln(2)}{R_{\star}}  
%
\label{eq_nh}
\end{eqnarray}

Inserting our assumed values into Eqn.\,\ref{eq_nh}, we derive a column density of $5 \times 10^{22}$\,cm$^{-2}$. In order to be consistent with our observed value of $N_{\mathrm{H}} = 10^{23}$\,cm$^{-2}$, we would need $\dot{M_{\star}}=2\times10^{-6}$\,$M_{\odot}$\,y$^{-1}$ which is a slightly larger, but still reasonable, stellar mass-loss rate. While this model can underestimate the true value of $N_{\mathrm{H}}$ for a given set of wind parameters, the derived column density should be higher if the system were viewed at superior conjunction (see Section\,\ref{subsec_eclipse}). A few caveats are in order. First, the intrinsic absorption in these sources can vary by an order of magnitude so a static model can not be expected to predict the full range of observed column densities. Second, the wind parameters that we assumed might not reflect the actual conditions of this system. Third, we relied on an orbital geometry that is simplistic (e.g. the inclination angle is zero) since we know little about the binary's orbit.

Iron fluorescence is another tell-tale sign of material near the high-energy emitter. An iron line was detected for this source in \xmm\ data \citep{Tom09}, and our \suz\ spectrum includes a possible iron line at 6.42$\pm$0.03\,keV with an upper limit on the equivalent width at $<$91\,eV, both of which are consistent with the corresponding values obtained with the \xmm\ spectrum ($E\left(\mathrm{K}\alpha\right)=$ 6.39$\pm$0.03\,keV, $EW=$42$\pm$12\,eV). The line is produced by continuum X-rays fluorescing Fe that is restricted to neutral (Fe\,I) or ionized up to Fe\,XV \citep[][and references therein]{Hou69,Nag89}. The line equivalent width combined with the column density are consistent with a model in which the fluorescence occurs within a spherically-symmetric shell of relatively cool matter \citep[][]{Nag89,Mat02,Kal04,Dra08}. Therefore, these measurements provide valuable clues to the properties of the matter around the CO.

The nature of the CO in \igr\ remains unknown. Periodic modulations attributable to a neutron star's spin have not yet been detected from this source despite long observations with \xmm\ \citep{Tom09} and now with \suz. One possible explanation for the non-detection of a pulsation is that the rotation and magnetic axes of the neutron star are nearly parallel. Another possibility is that an unfavorable geometric alignment causes the accretion beam to point in our direction throughout the rotation. As previously suggested by \citet{Tom09}, \igr\ might have a very slowly-rotating neutron star, e.g. $\sim$6\,ks such as in the magnetar candidate \object{IGR\,J16358$-$4726} \citep{Pat07}. In our data, a signal at this frequency is too close to the orbital period of \suz\ to be unambiguous. 

Instead of a neutron star, the CO might be a radio-quiet black hole candidate. The hard X-ray emission from a black hole candidate generally has a cutoff energy situated above 60\,keV. In \igr, the cutoff is at 20\,keV which suggests a neutron star primary \citep{Nag89}. Soft excess emission is often seen from accreting X-ray pulsars \citep{Hic04}, and this feature was detected in the \xmm\ spectrum of \igr\ \citep{Tom09}. There were residuals around 2\,keV from the optimal fitting model to the XIS spectrum, but the addition of a blackbody component did not improve the quality of the fit in a significant way.

Keep in mind that coherent pulsations have yet to be confirmed for many SGXBs including \linebreak \object{IGR\,J19140$+$0951}, which shares many of the same properties as \igr\ such as the companion's spectral type, a high intrinsic column density, and soft excess emission \citep[e.g.][]{Han07,Pra08,Tor09}. 
The lack of a measurable pulsation period or cyclotron absorption lines is common among SGXBs, and in particular among SFXTs, even though their spectral cutoff energies argue in favor of a NS primary.

\subsection{Is IGR\,J16207$-$5129 an eclipsing X-ray binary?}
\label{subsec_eclipse}

Our \suz\ observation ends with an epoch of $\sim30$\,ks in which the source has an average count rate in the soft and hard bands that is 5--10 times lower than before this epoch, and up to $\sim$20 times smaller at certain times. While SGXBs typically vary in luminosity by a factor 20 \citep{Wal06}, the long duration of the inactivity in this source is unusual and therefore intriguing. During a 12-ks stretch of O2b, the unabsorbed 0.5--10-keV flux is $4.3\times10^{-12}$\,\ergcms. This is lower than the lowest sustained flux (ks timescales) measured with \xmm: $7.0\times10^{-11}$\,\ergcms \citep{Tom09}. In addition, the epoch of flux minimum recorded by \xmm\ lasted $\sim$3\,ks \citep{Tom09}, which is 4 times shorter than the low-flux state that we observed with \suz. Note that the maximum flux that we measured is consistent with the flux measured during the high-intensity state from the \xmm\ observation: $\sim$$10^{-10}$\,\ergcms.

There are several mechanisms that could suppress the source count rate from \igr\ for a sustained period: an ``off'' state of the X-ray emitter; occulting clumps in the wind; eclipse of the shock region at the surface of the neutron star; an eclipse of the primary itself; and hydrodynamic effects. In the next few paragraphs, we will discuss each of these possibilities and the extent to which they are supported by the data.

The X-ray luminosity is driven by the accretion rate which is affected by, among other things, wind inhomogeneities such as clumpy material, by photoionization, and by the magnetic field of the CO. In general, persistent emission with stochastic variability is expected from such objects, but we point out that some SGXBs (e.g. Vela\,X-1), can enter short-lived inactive states \citep{Ino84}. The X-ray flux in this system decreases across all energies so varying absorption columns are not responsible. The ``off'' states are thought to be due to the passage of the NS in a cavity of low-density stellar wind combined with an onset of the propeller regime which temporarily inhibits accretion onto the compact object \citep{Kre08}. Typical timescales for these ``off'' states are on the order of a few hundred seconds, or at least an order of magnitude shorter than the period of inactivity that we observed in \igr.

The X-ray source might be temporarily shielded from us by optically-thick blobs of the supergiant star's wind passing through our line of sight. We would anticipate an increase in \nh\ on short timescales concurrent with a decrease in the X-ray flux, potentially followed by a flaring episode depending on the proximity of the blob to the CO, i.e. whether it is accreted. This process has been observed in SFXTs such as \object{IGR\,J17544$-$2619} \citep{Ram09}, as well as in ``classical'' SGXBs such as \object{Vela\,X-1} \citep{Kre08}. At least twice during O2a (at $t=12$ and $t=36$\,ks), the count rate from \igr\ remained low for several hundred seconds followed by flares and variable emission. These low states are also seen in the hard band (15--60\,keV). Combining the spectra from low-intensity states throughout O2, we did not detect a significant increase in absorption. Focusing solely on a 2.5-ks epoch beginning 12\,ks after the start of O2, the spectral fit yields \nh$=(19_{-6}^{+8})\times10^{22}$\,cm$^{-2}$, which is compatible with the \nh\ measured outside this epoch. Therefore, an occultation by a large wind clump along the line of sight to the compact object can not be responsible for the dip in luminosity since there is no accompanying increase in the \nh.

The reduction in intensity could be due to the orbital geometry of the system. \citet{Pra08} showed that in the inclined binary \object{IGR\,J19140$+$0951}, the \nh\ can be related to the orbital phase reaching a maximum at superior conjunction and a minimum at inferior conjunction. The hard X-ray flux (20--100\,keV), which is unaffected by photoelectric absorption, was actually suppressed at inferior conjunction. This is because the shock regions near the neutron star's surface, where high-velocity stellar winds encounter the ionized gas surrounding the CO's magnetosphere, are partially occulted by the CO at inferior conjunction. The minimum flux reported for \object{IGR\,J19140$+$0951} was listed as unabsorbed, so this occultation is unrelated to column density variations (or the lack thereof). On the other hand, if we are truly observing \igr\ at inferior conjunction, then, as in \object{IGR\,J19140$+$0951}, we should expect the column density at superior conjunction to be larger (by an order of magnitude) since the X-rays would have to travel through more obscuring material. This would imply \nh\ $\sim 10^{24}$\,cm$^{-2}$ at superior conjunction. This level of absorption has never been observed in \igr\ and it is only seen in extreme cases such as \object{IGR\,J16318$-$4848}. While we can not exclude that this effect is responsible for the diminished luminosity of \igr\ during O2b, we believe it is unrealistic given the implications for the column density at superior conjunction.


Magnetic and hydrodynamic effects (and possible photo-ionization) could be responsible for the apparent drop in luminosity \citep{Blo94}. Movement of the CO (and in the case of neutron stars, its ionizing magnetic field) through the accretion wind can disrupt and even inhibit the accretion of material leading to variability and even a temporary attenuation of the X-ray source. Such tidal-stream disruptions are accompanied by an increase in the column density of up to 20--30 times the average density of the wind, which we do not observe in \igr. 

The X-ray source may have been partially or fully-eclipsed by its supergiant companion star. We remark that a potential ingress phase is noticeable in the XIS light curve between MJD 54527.4 and 54527.6 (Fig.\,\ref{lc_x0_hxd}). During the eclipse of such systems, the continuum emission is believed to consist of a greater proportion of scattered X-rays. This should be visible as an evolving hardness ratio, and it is generally accompanied by an increase in the intensity of the iron line relative to the continuum emission \citep[e.g. \object{Vela\,X-1}:][and references therein]{Sat86}. Count rates in the soft and especially the hard bands were low during O2b, plaguing the hardness ratio of this epoch with large uncertainties, which made them difficult to compare with the hardness ratios from O2a.


\begin{figure}[!t] \centering
\includegraphics[width=8cm,angle=0]{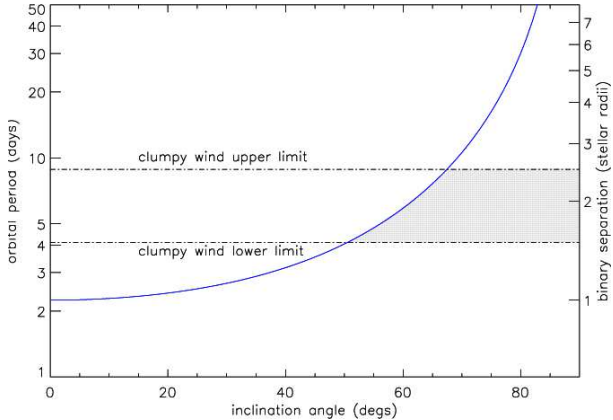}
\caption{Solutions to a model of an eclipsing SGXB \citep[Eq.\,\ref{eq_theta}:][]{Rap83} viewed under different inclination angles (degrees). The CO is assumed to be a neutron star of 1.4\,$M_{\odot}$, and the companion star's mass is set to 20\,$M_{\odot}$ with a radius of 20\,$R_{\odot}$. The eclipse is taken to last $\gtrsim$30\,ks (solid curve). The dot-dashed lines represent the limits of orbital radii of clumpy winds in SGXBs \citep{Neg08}. The shaded region shows the extent of the parameter space in which IGR\,J16207$-$5129 might be located. }
\label{orbit}
\end{figure}

In summary, the observed attenuation of the X-ray flux can not be explained by an increase in the column density since the measured absorption value does not change in a significant way between O2a and O2b, nor between the low and high-intensity states of O2. Finally, we mention that whatever mechanism is involved in driving the sporadic emission in SFXTs could be responsible for the prolonged dip that we see in \igr. The luminosity that we measured during the lowest states of O2b ($2\times10^{34}$\,\ergs) is similar to the quiescent flux reported for the SFXT \object{IGR\,J16479$-$4514} \citep{Wal07}. This is an order of magnitude greater than the quiescent luminosity expected from transient X-ray sources \citep[$10^{32-33}$\,\ergs: e.g.][]{Cam02}. Such long extinctions have never been reported for this object implying a high duty cycle that is consistent with its classification among SGXBs.

Hence, we are not able to provide any firm conclusions as to why the observed flux is consistently low for such a long time. Nevertheless, the eclipse interpretation remains a viable explanation and it allows us to consider the orbital configuration of the X-ray binary. An eclipse duration ($\Delta t$) of at least 30\,ks (0.35\,d) would set the lower limit on the orbital period at 0.7\,d. We do not see an egress phase in the light curve so 30\,ks should be considered as the lower limit on the eclipse duration. Given that a continuous 50-ks \xmm\ observation, which also includes periods of low activity (but notably, none are longer than $\sim$2\,ks), shows no evidence for an eclipse, the lower limit on the orbital period should be higher unless the CO is in an inclined orbit.

Following the procedure described in \citet{Boz08}, we can estimate the range of orbital periods ($P$) by using the fact that the separation of the centers of mass of the system ($a$) can be expressed as a function of the eclipse half-angle $\Theta_{\mathrm{e}}$ via \citep[][]{Rap83,Nag89}:

\begin{equation}
a = \frac { R_{\star} }{ \sqrt{ \cos^{2}i + \sin^{2}i \sin^{2} \Theta_{\mathrm{e}} }}   \,\,\,\,\,\,\mathrm{,}
\label{eq_theta}
\end{equation}

\noindent where $i$ is the inclination angle. For $\Theta_{\mathrm{e}}$, we take advantage of a simple trigonometric equivalence:

\begin{equation}
\frac { 2\Theta_{\mathrm{e}} }{ 2\pi } = \frac { \Delta t }{ P }
\end{equation}

We can then derive the orbital period (related to the binary separation $a$ through the generalized form of Kepler's Law) for various inclination angles. Figure\,\ref{orbit} presents the results of this model. Also shown are the clumpy winds that are expected in a radius range of 1.5 to 2.5\,$R_{\star}$ in SGXBs \citep{Neg08}. Based on our assumptions, we find that the parameter space is bounded by orbital periods between 4 and 9\,d with inclination angles $i \gtrsim 50^{\circ}$ (shaded region in Fig.\,\ref{orbit}). The full orbit is within the upper limit of the clumpy-wind radius ($a \lesssim 2.5 R_{\star}$), unless the orbital period is greater than $\sim$10\,d, which is unusual for SGXBs but plausible (e.g. \object{IGR\,J19140$+$0951} has an orbital period of $\sim$14\,d). For comparison, inclination angles for eclipsing HMXBs are generally between 60 and 85$^{\circ}$ with $\Theta_{\mathrm{e}} \sim 25^{\circ}$ \citep{Nag89}.

\section{Summary \& Conclusions}

Our \suz\ observation of \igr\ confirms the large intrinsic column density (\nh\ $=$\linebreak$(16.2_{-1.1}^{+0.9}) \times 10^{22}$\,cm$^{-2}$) that was reported previously, while demonstrating the long-term variability of the absorbing column. The broadband spectrum covered an energy range between 0.5 and 60\,keV with unprecedented spectral sensitivity above 15\,keV enabling us to constrain the cutoff energy for the first time. This cutoff is situated around $19_{-5}^{+6}$\,keV which is expected if the compact object is a neutron star, although we were not able to detect a coherent pulsation period or cyclotron lines. Intriguingly, the observation ends with a prolonged ($\sim30$\,ks) state of low activity where the luminosity decreases by a factor of 10 on average, and up to a factor of 20 between extremes, with little or no change in the absorbing column. While our data do not permit us to make firm conclusions to the nature of this event, one possibility is that the X-ray source is being eclipsed by its supergiant companion. 

Confirmation of the putative eclipse in \linebreak \igr\ requires further observations of the source. Only 4 IGRs are confirmed to be eclipsing binaries: \object{IGR\,J16418$-$4532}, \object{IGR\,J16479$-$4514}, \object{IGR\,J17252$-$3616}, and \object{IGR\,J18027$-$2016}. The first two are classified as SFXTs and the latter two are classical SGXBs. The demarcation between SFXTs and SGXBs is not as clear as originally believed, and a few systems (e.g. \object{IGR\,J16479$-$4514}, \object{IGR\,J18483$-$0311}, and \object{IGR\,J19294+1816}) might represent intermediate states between the SFXT and SGXB classes because of their peculiar properties. In this respect, the study of \igr\ will play an integral role in helping us understand the extent to which the emissivity differences between both populations stem from their unequal wind and orbital characteristics. 

\acknowledgments
The authors thank the anonymous referee whose detailed revision of the draft led to significant improvements in the quality of the manuscript. AB and JT acknowledge Suzaku Guest Observer Grant NNX08AB88G and INTEGRAL Guest Observer Grant NNX08AC91G. AB thanks Erica Kotta for a careful reading of the draft, and Nicolas Barri\`ere for help with IDL. This research has made use of: data obtained from the High Energy Astrophysics Science Archive Research Center (HEASARC) provided by NASA's Goddard Space Flight Center; the SIMBAD database operated at CDS, Strasbourg, France; NASA's Astrophysics Data System Bibliographic Services. Based on observations with INTEGRAL, an ESA project with instruments and science data center funded by ESA member states (especially the PI countries: Denmark, France, Germany, Italy, Switzerland, Spain), Poland and with the participation of Russia and the USA.

{\it Facilities:} \facility{INTEGRAL}, \facility{Suzaku}.

\bibliographystyle{apj}
\bibliography{16207.bib}

\begin{thebibliography}{55}
\expandafter\ifx\csname natexlab\endcsname\relax\def\natexlab#1{#1}\fi

\bibitem[{{Bird} {et~al.}(2004){Bird}, {Barlow}, {Bassani}, {Bazzano},
  {Bodaghee}, {Capitanio}, {Cocchi}, {Del Santo}, {Dean}, {Hill}, {Lebrun},
  {Malaguti}, {Malizia}, {Much}, {Shaw}, {Stephen}, {Terrier}, {Ubertini}, \&
  {Walter}}]{Bir04}
{Bird}, A.~J., {et~al.} 2004, \apjl, 607, L33

\bibitem[{{Bird} {et~al.}(2006){Bird}, {Barlow}, {Bassani}, {Bazzano},
  {B{\'e}langer}, {Bodaghee}, {Capitanio}, {Dean}, {Fiocchi}, {Hill}, {Lebrun},
  {Malizia}, {Mas-Hesse}, {Molina}, {Moran}, {Renaud}, {Sguera}, {Shaw},
  {Stephen}, {Terrier}, {Ubertini}, {Walter}, {Willis}, \& {Winkler}}]{Bir06}
---. 2006, \apj, 636, 765

\bibitem[{{Bird} {et~al.}(2007){Bird}, {Malizia}, {Bazzano}, {Barlow},
  {Bassani}, {Hill}, {B{\'e}langer}, {Capitanio}, {Clark}, {Dean}, {Fiocchi},
  {G{\"o}tz}, {Lebrun}, {Molina}, {Produit}, {Renaud}, {Sguera}, {Stephen},
  {Terrier}, {Ubertini}, {Walter}, {Winkler}, \& {Zurita}}]{Bir07}
---. 2007, \apjs, 170, 175

\bibitem[{{Bird} {et~al.}(2010){Bird}, {Bazzano}, {Bassani}, {Capitanio},
  {Fiocchi}, {Hill}, {Malizia}, {McBride}, {Scaringi}, {Sguera}, {Stephen},
  {Ubertini}, {Dean}, {Lebrun}, {Terrier}, {Renaud}, {Mattana}, {G{\"o}tz},
  {Rodriguez}, {Belanger}, {Walter}, \& {Winkler}}]{Bir10}
---. 2010, \apjs, 186, 1

\bibitem[{{Blondin}(1994)}]{Blo94}
{Blondin}, J.~M. 1994, \apj, 435, 756

\bibitem[{{Bodaghee} {et~al.}(2006){Bodaghee}, {Walter}, {Zurita Heras},
  {Bird}, {Courvoisier}, {Malizia}, {Terrier}, \& {Ubertini}}]{Bod06}
{Bodaghee}, A., {Walter}, R., {Zurita Heras}, J.~A., {Bird}, A.~J.,
  {Courvoisier}, T., {Malizia}, A., {Terrier}, R., \& {Ubertini}, P. 2006,
  \aap, 447, 1027

\bibitem[{{Bodaghee} {et~al.}(2007){Bodaghee}, {Courvoisier}, {Rodriguez},
  {Beckmann}, {Produit}, {Hannikainen}, {Kuulkers}, {Willis}, \&
  {Wendt}}]{Bod07}
{Bodaghee}, A., {et~al.} 2007, \aap, 467, 585

\bibitem[{{Bozzo} {et~al.}(2008){Bozzo}, {Stella}, {Israel}, {Falanga}, \&
  {Campana}}]{Boz08}
{Bozzo}, E., {Stella}, L., {Israel}, G., {Falanga}, M., \& {Campana}, S. 2008,
  \mnras, 391, L108

\bibitem[{{Campana} {et~al.}(2002){Campana}, {Stella}, {Israel}, {Moretti},
  {Parmar}, \& {Orlandini}}]{Cam02}
{Campana}, S., {Stella}, L., {Israel}, G.~L., {Moretti}, A., {Parmar}, A.~N.,
  \& {Orlandini}, M. 2002, \apj, 580, 389

\bibitem[{{Castor} {et~al.}(1975){Castor}, {Abbott}, \& {Klein}}]{Cas75}
{Castor}, J.~I., {Abbott}, D.~C., \& {Klein}, R.~I. 1975, \apj, 195, 157

\bibitem[{{Crowther} {et~al.}(2006){Crowther}, {Lennon}, \& {Walborn}}]{Cro06}
{Crowther}, P.~A., {Lennon}, D.~J., \& {Walborn}, N.~R. 2006, \aap, 446, 279

\bibitem[{{Dickey} \& {Lockman}(1990)}]{Dic90}
{Dickey}, J.~M., \& {Lockman}, F.~J. 1990, \araa, 28, 215

\bibitem[{{Drake} {et~al.}(2008){Drake}, {Ercolano}, \& {Swartz}}]{Dra08}
{Drake}, J.~J., {Ercolano}, B., \& {Swartz}, D.~A. 2008, \apj, 678, 385

\bibitem[{{Filliatre} \& {Chaty}(2004)}]{Fil04}
{Filliatre}, P., \& {Chaty}, S. 2004, \apj, 616, 469

\bibitem[{{Hannikainen} {et~al.}(2007){Hannikainen}, {Rawlings}, {Muhli},
  {Vilhu}, {Schultz}, \& {Rodriguez}}]{Han07}
{Hannikainen}, D.~C., {Rawlings}, M.~G., {Muhli}, P., {Vilhu}, O., {Schultz},
  J., \& {Rodriguez}, J. 2007, \mnras, 380, 665

\bibitem[{{Hickox} {et~al.}(2004){Hickox}, {Narayan}, \& {Kallman}}]{Hic04}
{Hickox}, R.~C., {Narayan}, R., \& {Kallman}, T.~R. 2004, \apj, 614, 881

\bibitem[{{House}(1969)}]{Hou69}
{House}, L.~L. 1969, \apjs, 18, 21

\bibitem[{{Inoue} {et~al.}(1984){Inoue}, {Ogawara}, {Waki}, {Ohashi},
  {Hayakawa}, {Kunieda}, {Nagase}, \& {Tsunemi}}]{Ino84}
{Inoue}, H., {Ogawara}, Y., {Waki}, I., {Ohashi}, T., {Hayakawa}, S.,
  {Kunieda}, H., {Nagase}, F., \& {Tsunemi}, H. 1984, \pasj, 36, 709

\bibitem[{{Kalberla} {et~al.}(2005){Kalberla}, {Burton}, {Hartmann}, {Arnal},
  {Bajaja}, {Morras}, \& {P{\"o}ppel}}]{Kal05}
{Kalberla}, P.~M.~W., {Burton}, W.~B., {Hartmann}, D., {Arnal}, E.~M.,
  {Bajaja}, E., {Morras}, R., \& {P{\"o}ppel}, W.~G.~L. 2005, \aap, 440, 775

\bibitem[{{Kallman} {et~al.}(2004){Kallman}, {Palmeri}, {Bautista}, {Mendoza},
  \& {Krolik}}]{Kal04}
{Kallman}, T.~R., {Palmeri}, P., {Bautista}, M.~A., {Mendoza}, C., \& {Krolik},
  J.~H. 2004, \apjs, 155, 675

\bibitem[{{Koyama} {et~al.}(2007){Koyama}, {Tsunemi}, {Dotani}, {Bautz},
  {Hayashida}, {Tsuru}, {Matsumoto}, {Ogawara}, {Ricker}, {Doty}, {Kissel},
  {Foster}, {Nakajima}, {Yamaguchi}, {Mori}, {Sakano}, {Hamaguchi},
  {Nishiuchi}, {Miyata}, {Torii}, {Namiki}, {Katsuda}, {Matsuura}, {Miyauchi},
  {Anabuki}, {Tawa}, {Ozaki}, {Murakami}, {Maeda}, {Ichikawa}, {Prigozhin},
  {Boughan}, {Lamarr}, {Miller}, {Burke}, {Gregory}, {Pillsbury}, {Bamba},
  {Hiraga}, {Senda}, {Katayama}, {Kitamoto}, {Tsujimoto}, {Kohmura}, {Tsuboi},
  \& {Awaki}}]{Koy07}
{Koyama}, K., {et~al.} 2007, \pasj, 59, 23

\bibitem[{{Kreykenbohm} {et~al.}(2008){Kreykenbohm}, {Wilms}, {Kretschmar},
  {Torrej{\'o}n}, {Pottschmidt}, {Hanke}, {Santangelo}, {Ferrigno}, \&
  {Staubert}}]{Kre08}
{Kreykenbohm}, I., {et~al.} 2008, \aap, 492, 511

\bibitem[{{Krivonos} {et~al.}(2007){Krivonos}, {Revnivtsev}, {Lutovinov},
  {Sazonov}, {Churazov}, \& {Sunyaev}}]{Kri07}
{Krivonos}, R., {Revnivtsev}, M., {Lutovinov}, A., {Sazonov}, S., {Churazov},
  E., \& {Sunyaev}, R. 2007, \aap, 475, 775

\bibitem[{{Lebrun} {et~al.}(2003){Lebrun}, {Leray}, {Lavocat}, {Cr{\' e}tolle},
  {Arqu{\` e}s}, {Blondel}, {Bonnin}, {Bou{\` e}re}, {Cara}, {Chaleil}, {Daly},
  {Desages}, {Dzitko}, {Horeau}, {Laurent}, {Limousin}, {Mathy}, {Mauguen},
  {Meignier}, {Molini{\' e}}, {Poindron}, {Rouger}, {Sauvageon}, \&
  {Tourrette}}]{Leb03}
{Lebrun}, F., {et~al.} 2003, \aap, 411, L141

\bibitem[{{Levine} {et~al.}(2004){Levine}, {Rappaport}, {Remillard}, \&
  {Savcheva}}]{Lev04}
{Levine}, A.~M., {Rappaport}, S., {Remillard}, R., \& {Savcheva}, A. 2004,
  \apj, 617, 1284

\bibitem[{{Liu} {et~al.}(2006){Liu}, {van Paradijs}, \& {van den
  Heuvel}}]{Liu06}
{Liu}, Q.~Z., {van Paradijs}, J., \& {van den Heuvel}, E.~P.~J. 2006, \aap,
  455, 1165

\bibitem[{{Masetti} {et~al.}(2006){Masetti}, {Morelli}, {Palazzi}, {Galaz},
  {Bassani}, {Bazzano}, {Bird}, {Dean}, {Israel}, {Landi}, {Malizia},
  {Minniti}, {Schiavone}, {Stephen}, {Ubertini}, \& {Walter}}]{Mas06c}
{Masetti}, N., {et~al.} 2006, \aap, 459, 21

\bibitem[{{Matt}(2002)}]{Mat02}
{Matt}, G. 2002, \mnras, 337, 147

\bibitem[{{Morris} {et~al.}(2009){Morris}, {Smith}, {Markwardt}, {Mushotzky},
  {Tueller}, {Kallman}, \& {Dhuga}}]{Mor09}
{Morris}, D.~C., {Smith}, R.~K., {Markwardt}, C.~B., {Mushotzky}, R.~F.,
  {Tueller}, J., {Kallman}, T.~R., \& {Dhuga}, K.~S. 2009, \apj, 699, 892

\bibitem[{{Nagase}(1989)}]{Nag89}
{Nagase}, F. 1989, \pasj, 41, 1

\bibitem[{{Negueruela} \& {Schurch}(2007)}]{Neg07}
{Negueruela}, I., \& {Schurch}, M.~P.~E. 2007, \aap, 461, 631

\bibitem[{{Negueruela} {et~al.}(2006){Negueruela}, {Smith}, {Reig}, {Chaty}, \&
  {Torrej{\'o}n}}]{Neg06}
{Negueruela}, I., {Smith}, D.~M., {Reig}, P., {Chaty}, S., \& {Torrej{\'o}n},
  J.~M. 2006, in ESA Special Publication, Vol. 604, The X-ray Universe 2005,
  ed. {A.~Wilson}, 165--+

\bibitem[{{Negueruela} {et~al.}(2008){Negueruela}, {Torrej{\'o}n}, {Reig},
  {Rib{\'o}}, \& {Smith}}]{Neg08}
{Negueruela}, I., {Torrej{\'o}n}, J.~M., {Reig}, P., {Rib{\'o}}, M., \&
  {Smith}, D.~M. 2008, in American Institute of Physics Conference Series, Vol.
  1010, A Population Explosion: The Nature \& Evolution of X-ray Binaries in
  Diverse Environments, ed. {R.~M.~Bandyopadhyay, S.~Wachter, D.~Gelino, \&
  C.~R.~Gelino}, 252--256

\bibitem[{{Nespoli} {et~al.}(2008){Nespoli}, {Fabregat}, \&
  {Mennickent}}]{Nes08}
{Nespoli}, E., {Fabregat}, J., \& {Mennickent}, R.~E. 2008, \aap, 486, 911

\bibitem[{{Paizis} {et~al.}(2009){Paizis}, {Ebisawa}, {Takahashi}, {Dotani},
  {Kohmura}, {Kokubun}, {Rodriguez}, {Ueda}, {Walter}, {Yamada}, {Yamaoka}, \&
  {Yuasa}}]{Pai09}
{Paizis}, A., {et~al.} 2009, \pasj, 61, 107

\bibitem[{{Patel} {et~al.}(2007){Patel}, {Zurita}, {Del Santo}, {Finger},
  {Kouveliotou}, {Eichler}, {G{\"o}{\u g}{\"u}{\c s}}, {Ubertini}, {Walter},
  {Woods}, {Wilson}, {Wachter}, \& {Bazzano}}]{Pat07}
{Patel}, S.~K., {et~al.} 2007, \apj, 657, 994

\bibitem[{{Prat} {et~al.}(2008){Prat}, {Rodriguez}, {Hannikainen}, \&
  {Shaw}}]{Pra08}
{Prat}, L., {Rodriguez}, J., {Hannikainen}, D.~C., \& {Shaw}, S.~E. 2008,
  \mnras, 389, 301

\bibitem[{{Rahoui} {et~al.}(2008){Rahoui}, {Chaty}, {Lagage}, \&
  {Pantin}}]{Rah08}
{Rahoui}, F., {Chaty}, S., {Lagage}, P., \& {Pantin}, E. 2008, \aap, 484, 801

\bibitem[{{Rampy} {et~al.}(2009){Rampy}, {Smith}, \& {Negueruela}}]{Ram09}
{Rampy}, R.~A., {Smith}, D.~M., \& {Negueruela}, I. 2009, \apj, 707, 243

\bibitem[{{Rappaport} \& {Joss}(1983)}]{Rap83}
{Rappaport}, S.~A., \& {Joss}, P.~C. 1983, in Accretion-Driven Stellar X-ray
  Sources, ed. {W.~H.~G.~Lewin \& E.~P.~J.~van den Heuvel}, 1--39

\bibitem[{{Robba} {et~al.}(2001){Robba}, {Burderi}, {Di Salvo}, {Iaria}, \&
  {Cusumano}}]{Rob01}
{Robba}, N.~R., {Burderi}, L., {Di Salvo}, T., {Iaria}, R., \& {Cusumano}, G.
  2001, \apj, 562, 950

\bibitem[{{Rodriguez} {et~al.}(2010){Rodriguez}, {Tomsick}, \&
  {Bodaghee}}]{Rod10}
{Rodriguez}, J., {Tomsick}, J.~A., \& {Bodaghee}, A. 2010, ArXiv e-prints

\bibitem[{{Sato} {et~al.}(1986){Sato}, {Hayakawa}, \& {Nagase}}]{Sat86}
{Sato}, N., {Hayakawa}, S., \& {Nagase}, F. 1986, \apss, 119, 81

\bibitem[{{Sguera} {et~al.}(2006){Sguera}, {Bazzano}, {Bird}, {Dean},
  {Ubertini}, {Barlow}, {Bassani}, {Clark}, {Hill}, {Malizia}, {Molina}, \&
  {Stephen}}]{Sgu06}
{Sguera}, V., {et~al.} 2006, \apj, 646, 452

\bibitem[{{Sidoli} {et~al.}(2008){Sidoli}, {Romano}, {Mangano}, {Pellizzoni},
  {Kennea}, {Cusumano}, {Vercellone}, {Paizis}, {Burrows}, \&
  {Gehrels}}]{Sid08}
{Sidoli}, L., {et~al.} 2008, \apj, 687, 1230

\bibitem[{{Takahashi} {et~al.}(2007){Takahashi}, {Abe}, {Endo}, {Endo}, {Ezoe},
  {Fukazawa}, {Hamaya}, {Hirakuri}, {Hong}, {Horii}, {Inoue}, {Isobe}, {Itoh},
  {Iyomoto}, {Kamae}, {Kasama}, {Kataoka}, {Kato}, {Kawaharada}, {Kawano},
  {Kawashima}, {Kawasoe}, {Kishishita}, {Kitaguchi}, {Kobayashi}, {Kokubun},
  {Kotoku}, {Kouda}, {Kubota}, {Kuroda}, {Madejski}, {Makishima}, {Masukawa},
  {Matsumoto}, {Mitani}, {Miyawaki}, {Mizuno}, {Mori}, {Mori}, {Murashima},
  {Murakami}, {Nakazawa}, {Niko}, {Nomachi}, {Okada}, {Ohno}, {Oonuki}, {Ota},
  {Ozawa}, {Sato}, {Shinoda}, {Sugiho}, {Suzuki}, {Taguchi}, {Takahashi},
  {Takahashi}, {Takeda}, {Tamura}, {Tamura}, {Tanaka}, {Tanihata}, {Tashiro},
  {Terada}, {Tominaga}, {Uchiyama}, {Watanabe}, {Yamaoka}, {Yanagida}, \&
  {Yonetoku}}]{Tak07}
{Takahashi}, T., {et~al.} 2007, \pasj, 59, 35

\bibitem[{{Tomsick} {et~al.}(2006){Tomsick}, {Chaty}, {Rodriguez}, {Foschini},
  {Walter}, \& {Kaaret}}]{Tom06}
{Tomsick}, J.~A., {Chaty}, S., {Rodriguez}, J., {Foschini}, L., {Walter}, R.,
  \& {Kaaret}, P. 2006, \apj, 647, 1309

\bibitem[{{Tomsick} {et~al.}(2009){Tomsick}, {Chaty}, {Rodriguez}, {Walter},
  {Kaaret}, \& {Tovmassian}}]{Tom09}
{Tomsick}, J.~A., {Chaty}, S., {Rodriguez}, J., {Walter}, R., {Kaaret}, P., \&
  {Tovmassian}, G. 2009, \apj, 694, 344

\bibitem[{{Tomsick} {et~al.}(2004){Tomsick}, {Lingenfelter}, {Corbel},
  {Goldwurm}, \& {Kaaret}}]{Tom04}
{Tomsick}, J.~A., {Lingenfelter}, R., {Corbel}, S., {Goldwurm}, A., \&
  {Kaaret}, P. 2004, in ESA Special Publication, Vol. 552, 5th INTEGRAL
  Workshop on the INTEGRAL Universe, ed. {V.~Schoenfelder, G.~Lichti, \&
  C.~Winkler}, 413--416

\bibitem[{{Torrej{\'o}n} {et~al.}(2009){Torrej{\'o}n}, {Negueruela}, {Smith},
  \& {Harrison}}]{Tor09}
{Torrej{\'o}n}, J.~M., {Negueruela}, I., {Smith}, D.~M., \& {Harrison}, T.~E.
  2009, ArXiv e-prints

\bibitem[{{Walter} \& {Zurita Heras}(2007)}]{Wal07}
{Walter}, R., \& {Zurita Heras}, J. 2007, \aap, 476, 335

\bibitem[{{Walter} {et~al.}(2004){Walter}, {Bodaghee}, {Barlow}, {Bird},
  {Dean}, {Hill}, {Shaw}, {Bazzano}, {Ubertini}, {Bassani}, {Malizia},
  {Stephen}, {Belanger}, {Lebrun}, \& {Terrier}}]{Wal04}
{Walter}, R., {et~al.} 2004, The Astronomer's Telegram, 229, 1

\bibitem[{{Walter} {et~al.}(2006){Walter}, {Zurita Heras}, {Bassani},
  {Bazzano}, {Bodaghee}, {Dean}, {Dubath}, {Parmar}, {Renaud}, \&
  {Ubertini}}]{Wal06}
---. 2006, \aap, 453, 133

\bibitem[{{Wilms} {et~al.}(2000){Wilms}, {Allen}, \& {McCray}}]{Wil00}
{Wilms}, J., {Allen}, A., \& {McCray}, R. 2000, \apj, 542, 914

\bibitem[{{Winkler} {et~al.}(2003){Winkler}, {Courvoisier}, {Di Cocco},
  {Gehrels}, {Gim{\'e}nez}, {Grebenev}, {Hermsen}, {Mas-Hesse}, {Lebrun},
  {Lund}, {Palumbo}, {Paul}, {Roques}, {Schnopper}, {Sch{\"o}nfelder},
  {Sunyaev}, {Teegarden}, {Ubertini}, {Vedrenne}, \& {Dean}}]{Win03}
{Winkler}, C., {et~al.} 2003, \aap, 411, L1

\end{thebibliography}

\clearpage

\end{document}